\documentclass[10pt]{article}
\usepackage{amsmath,amsfonts}
\usepackage{epsfig}
\usepackage{comment}

\newcommand{\smallspace}{\mskip 2mu minus 1mu}
\newcommand{\tinyspace}{\mskip 1mu}

\def\squareforqed{\hbox{\rlap{$\sqcap$}$\sqcup$}}
\def\qed{\ifmmode\squareforqed\else{\unskip\nobreak\hfil
\penalty50\hskip1em\null\nobreak\hfil\squareforqed
\parfillskip=0pt\finalhyphendemerits=0\endgraf}\fi}

\newtheorem{theorem}{Theorem}
\newtheorem{lemma}[theorem]{Lemma}
\newenvironment{proof}{\begin{trivlist}\item[]{\flushleft\bf Proof }}
{\qed\end{trivlist}}
\newenvironment{proofof}[1]{\begin{trivlist}
\item[]{\flushleft\bf Proof of #1 }}
{\qed\end{trivlist}}

\newcommand{\ket}[1]{\mbox{$| #1 \rangle$}}
\newcommand{\iket}[2]{\mbox{$| #1$--$#2\rangle$}}
\newcommand{\bra}[1]{\mbox{$\langle #1 |$}}

\newcommand{\inner}[2]{\mbox{$\langle #1 | #2 \rangle$}}
\newcommand{\tinner}[3]{\mbox{$\langle #1 |#2| #3 \rangle$}}

\newcommand{\weight}{\ensuremath{\omega}}

\newcommand{\nth}[1]{$#1$th}
\newcommand{\emm}{m}
\newcommand{\uell}{\ell}
\newcommand{\up}[2]{\mbox{$#1^{(#2)}$}}

\newcommand{\complex}{{\mathbb C}}

\newcommand{\op}[1]{{\mathsf{#1}}}
\newcommand{\identity}{\mbox{\boldmath$\tinyspace{\mathsf{I}}\tinyspace$}}
\newcommand{\projection}{\op{P}}

\newcommand{\betastar}{\beta^{*}}
\newcommand{\betastaralign}{\beta^{\vphantom{*}}}

\title{Bounds on quantum ordered searching}

\author{Peter H{\o}yer$\smallspace$\thanks{$\smallspace$BRICS, 
University of Aarhus, \mbox{DK--8000} {\AA}rhus~C, Denmark.  
email:~\mbox{\texttt{hoyer}\textbf{\char"40}\texttt{brics.dk}}.
BRICS, Basic Research in Computer Science,
is a Centre of the Danish National Research Foundation.}
\and \addtocounter{footnote}{1}
Jan 
Neerbek$\smallspace$\thanks{$\smallspace$Department 
of Computer Science, University of Aarhus, 
\mbox{DK--8000} {\AA}rhus~C, Denmark.  
email:~\mbox{\texttt{neerbek}\textbf{\char"40}\texttt{daimi.au.dk}}.}}

\date{}

\begin{document}
\maketitle

\begin{abstract}
We~prove that any exact quantum algorithm searching an ordered list 
of $N$~elements requires more than $\frac{1}{\pi}(\ln(N)-1)$ 
queries to the list. 
This improves upon the previously best known lower bound 
of $\frac{1}{12}\log_2(N) - O(1)$.
Our proof is based on a weighted all-pairs inner product argument,
and it generalizes to bounded-error quantum algorithms.

The currently best known upper bound for exact searching is 
roughly $0.526 \log_2(N)$.
We~give an exact quantum algorithm that 
uses $\log_3(N) + O(1)$ queries,
which is roughly $0.631 \log_2(N)$.
The main principles in our algorithm are an quantum parallel use of the
classical binary search algorithm and a method that allows basis
states in superpositions to communicate.

{\ }

\end{abstract}


\section{Introduction}
\label{sec:intro}

Searching ordered lists is one of the most fundamental and most 
studied problems in the theory of algorithms.
Given an ordered list and some element, the ordered search problem 
is to find the index of that element.
We~are interested the quantum black~box complexity of searching,
which is a quantum analogue of the decision tree complexity.

Formally, the input is a sorted Boolean list of size~$N$ and 
the problem is to find the index of the leftmost~1.
We~assume that the list is given as a black~box, so the only
way we can obtain information about the list is via queries.
The input to a query is an index and the output is the bit
of the list at that index.  
We~assume that not all values of the
list are~0 and hence the problem is always well-defined.

The classical query complexity of this problem is,
as is well known, exactly $\lceil \log_2(N)\rceil$.
Farhi, Goldstone, Gutmann and Sipser~\cite{FGGS2} have shown that 
on a quantum computer, we can solve this problem using only 
roughly $0.526 \log_2(N)$ queries.
The previously best known lower bound is due to 
Ambainis~\cite{Ambainis1} who recently proved that 
$\frac{1}{12} \log_2(N) - O(1)$ queries are required.
We~improve this lower bound to $\frac{1}{\pi}(\ln(N)-1)$
which is about $0.221 \log_2(N)$.
Thus, a speedup by a constant factor somewhere between 
$1.9$ and $4.6$ is achievable.

There are at least three reasons why the quantum complexity
of the ordered searching problem seems to be of interest.
Firstly because of its significance in algorithmics in general.
Secondly because the problem possesses some symmetries and 
periodicities of a different nature than previously studied problems
in quantum algorithmics.
Determining symmetries and periodicities seems to be a primary
ability of quantum computers and it is not at all clear 
how far-reaching this skill~is.
Thirdly because ordered searching represents
a non-Boolean non-symmetric function.
A~(partial) function is said to be symmetric if for each 
possible Hamming weight,
the function is either not defined for any input of that
Hamming weight, or it takes the same
value on all inputs of that Hamming weight.
Searching ordered lists is not such a problem since it is
only defined for exactly~1 input of each Hamming weight.
Only few non-trivial quantum bounds for non-Boolean 
non-symmetric functions are known.

We~prove our lower bound of $\frac{1}{\pi}(\ln(N)-1)$
by utilizing what we refer to as 
a weighted all-pairs inner product argument.
We~hope that this proof technique, which extends previous 
work by especially 
Bennett, Bernstein, Brassard and Vazirani~\cite{BBBV}
and Ambainis~\cite{Ambainis2}, will be of use elsewhere.
We~first give a description of our proof technique
in~Section~\ref{sec:blackbox}, by utilizing which we prove
a general lower bound in Section~\ref{sec:general}.
We~then apply the technique to ordered searching 
in Section~\ref{sec:search}.

In~Section~\ref{sec:upper}, 
we give a new quantum algorithm for ordered searching.
Our algorithm is exact and uses $\log_3(N) + O(1)$ queries,
which is roughly $0.631 \log_2(N)$. 
This is better than classically, but not as good as the 
currently best known algorithm which requires only roughly
$0.526 \log_2(N)$ queries~\cite{FGGS2}.
Whereas most quantum algorithms is based on Fourier transform and
amplitude amplification, our algorithm can be viewed as a quantum 
version of the classical (binary search) algorithm.
Our main idea is to run several applications of the binary
search algorithm in quantum parallel, and let them find the
element in teamwork.
We let some applications concentrate primarily on some parts of 
the inputs, meanwhile other applications concentrate primarily on 
other parts of the input.
Depending on where the searched element is, this implies
that some applications learn a great deal (about where the 
element is), whereas others learn only little.
We~then, as our main trick, use a method for letting the
different applications learn from each other, in the sense that
the applications that know the most about the input inform the 
other applications, hereby letting everybody learn a great deal
about the input.  Letting the different applications 
work as a team allows us reduce the complexity from $\log_2(N)$
to roughly $\log_3(N)$.

Finally, we conclude in Section~\ref{sec:conclusion} with
some remarks and open questions.

\section{Quantum black~box computing}
\label{sec:blackbox}
We~use the so-called black~box model in which the input is 
given as an oracle and 
our measure of complexity is the number of queries to the 
oracle~\cite{Berthiaume,BBCMdW}.
Fix some positive integer $N>0$.  
The input $x =(x_0,\ldots,x_{N-1})\in \{0,1\}^N$ 
is given as an oracle, and the only way we can access the 
bits of the oracle is via queries.  
A~query implements the operator
\begin{equation}\label{eq:oracle}
\op{O}_x: \quad \ket{z;i} \;\longmapsto\; 
\begin{cases}
(-1)^{x_i} \ket{z;i} & \text{if $0\leq i < N$}\\
\hphantom{(-1)^{x_i}} \ket{z;i} & \text{if $i \geq N$.}
\end{cases}
\end{equation}
Here $i$ and $z$ are non-negative integers.
By~a query to oracle~$x$ we mean an application of the 
unitary operator~$\op{O}_x$.  
We~sometimes refer to~$\op{O}_x$ as the oracle.
A~quantum algorithm~$A$ that uses $T$ queries to an 
oracle~$\op{O}$ is a unitary operator of the form
\begin{equation}\label{eq:algorithm}
A = (\op{U} \op{O})^T \op{U}.
\end{equation}
We~always apply algorithm~$A$ on the initial state~$\ket{0}$,
and after applying~$A$, we always measure the final 
state in the computational basis.
Thus, a quantum algorithm for oracle quantum computing
is defined by specifying a unitary operator~$\op{U}$ and
a number of iterations~$T$.  
Our model for oracle quantum computing is slightly
different from, but equivalent to,
the ``standard'' model used for example in~\cite{BBCMdW}.
We~favor utilizing this model, since hereby oracle~$\op{O}_x$ 
is a diagonal matrix with respect to the computational basis.

Consider the computation of some function 
$f:\{0,1\}^N \rightarrow \{0,1\}^\emm$.
After applying quantum algorithm~$A$ on~$\ket{0}$, we measure 
the $\emm$ rightmost qubits of~$A\ket{0}$ and output the outcome~$w$.
The success probability $p_x$ of~$A$ on input~$x \in \{0,1\}^N$ 
is defined as the probability that $w=f(x)$.
For complete functions $f:\{0,1\}^N \rightarrow \{0,1\}^\emm$,
we define the success probability of~$A$ as the minimum
of $p_x$ over all $x \in \{0,1\}^N$.
For partial functions $f:S \rightarrow \{0,1\}^\emm$, 
where $S \subseteq \{0,1\}^N$, we take the minimum over~$S$ only.

\subsection{Distinguishing oracles}
\label{subsec:distinguish}

There is a key conceptual idea which is used in this paper
and which we would like to emphasize, and that is that we are
concerned about \emph{distinguishing} oracles rather than
\emph{determining} the value of the function~$f$.
This is a line of thought that we have learned primarily from
Ettinger~\cite{Ettinger}
and Farhi, Goldstone, Gutmann and Sipser~\cite{FGGS3}.
Thus the general question we are interested in is that of 
distinguishing operators 
as opposed to distinguishing states or computing functions.
We~now give our version of such an approach applied to 
black~box computing.

The basic observation is that some oracles are easier 
to distinguish between than others.
Intuitively, it is easier to distinguish between the 
oracle $x = (0,0,\ldots,0)$ of all zeros and the oracle
$x = (1,1,\ldots,1)$ of all ones, than it is to 
distinguish between two oracles of almost the same Hamming 
weight.  One reason for a problem to be hard, is
if we want to distinguish between oracles of almost the same 
Hamming weight. 
In~general, if~a problem is hard,
it is because that there are some pairs of oracles
that are hard to distinguish.  Other pairs of oracles might
be easy to distinguish, but that may not necessarily
lower the complexity of the problem.

For a given problem we therefore want 
to identify the \emph{pairs of oracles} that are hard 
to \emph{distinguish}.  
This idea is applicable when proving lower as well as upper bounds.
{To}~capture the hardness of distinguishing each pair of oracles,
we introduce a \emph{weight function}
\begin{equation}
\weight: \quad \{0,1\}^N \times \{0,1\}^N \rightarrow \Re_{+}
\end{equation}
that takes non-negative real values.
The harder an oracle~$x$ is to distinguish from 
an oracle~$y$, the more weight we put on the pair~$(x,y)$.
The total weight~$W$ distributed is the 
sum of $\weight(x,y)$ over all pairs 
${(x,y) \in \{0,1\}^N \times \{0,1\}^N}$.
We~do not want to put any restrictions on~$\weight$ in general,
though for many applications we probably want $\weight$ 
to be symmetric, normalized and take the value~0 along the
diagonal.

The weight function allows us to easily capture any complete
as well as partial function.
Let $f: S \times S \rightarrow \{0,1\}^\emm$ be a function
of interest, where $S \subseteq \{0,1\}^N$.
We~say that \emph{$\weight$ is a weight function for~$f$}
if whenever $f(x)=f(y)$ then $\weight(x,y)=0$, 
and if for every pair $(x,y) \not\in S \times S$
we have $\weight(x,y)=0$.
Hereby, we may ignore~$f$ and just consider the scenario
in which we are given weight function~$\weight$.

\section{General lower bound}
\label{sec:general}
The first general technique for proving lower bounds for
quantum computing was introduced by Bennett, Bernstein,
Brassard and Vazirani in their influential paper~\cite{BBBV}.
Their beautiful technique is nicely described in Vazirani's 
exposition~\cite{Vazirani}.
Our technique is a natural generalization of theirs, but it 
can also be viewed as a generalization of 
Ambainis' powerful entanglement lower bound approach 
recently proposed in~\cite{Ambainis2} 
(provided one casts his technique using a language 
similar to the one used in~\cite{BBBV} and here).

Here is the basic idea:
Consider a quantum algorithm $A = (\op{U}\op{O})^T \op{U}$ 
that we use to
distinguish between two oracles $x,y \in \{0,1\}^N$.
Our initial state is~$\ket{0}$.
After $j$~iterations, our state is 
$\ket{\psi_x^j} = (\op{U}\op{O}_x)^j \op{U}\ket{0}$ 
if we are given oracle~$x$, 
and it is $\ket{\psi_y^j} = (\op{U}\op{O}_y)^j \op{U}\ket{0}$ 
if we are given oracle~$y$.
Two quantum states are distinguishable with high probability 
if and only if they are almost orthogonal.
If~the states $\ket{\psi_x^j}$ and $\ket{\psi_y^j}$ have large
overlap, then they cannot be distinguished with high probability,
and hence more queries are required.
If~a query can separate two states
$\ket{\psi_x^j}$ and~$\ket{\psi_y^j}$ by only a small additional 
amount, then many queries are required.  

We~have to choose how to measure the overlap of states among
the plentiful studied measures.  
We~pick here the probably most simple possibility: inner products.
Two states can be distinguished with certainty
if and only if their inner product is zero.
Furthermore, two states can be distinguished with high probability
if and only if their inner product is of small absolute value.

\begin{lemma}\label{lm:epsilon}
Suppose that we are given one of two states
 $\ket{\Psi_x}, \ket{\Psi_y}$.
There exists some measurement that will correctly
determine which of the two states we are given 
with error probability at most~$\epsilon$ if and only if
$|\inner{\Psi_x}{\Psi_y}| \leq 2 \sqrt{\epsilon(1-\epsilon)}$.
\end{lemma}

We~are not only interested in distinguishing two particular
oracles, but many oracles, and thus we will use an ``all-pairs inner 
product'' measure.
But as we discussed in the previous section, some oracles
are harder to distinguish than others, and this leads us to
our final choice: we use an \emph{all-pairs inner product 
measure weighted by~$\weight$}.
We~now formalize this approach.

Let $A = (\op{U}\op{O})^T \op{U}$ be any quantum algorithm.
For every oracle $x \in \{0,1\}^N$ and every integer $j \geq 0$, let
\begin{equation}
\ket{\psi_x^j} \smallspace=\smallspace (\op{U}\op{O}_x)^j \op{U}
\smallspace\ket{0}
\end{equation}
denote the state of the computer after applying $j$ iterations
using oracle~$\op{O}_x$. 
For every integer $j\geq 0$, let
\begin{equation}
W_j =
 \sum_{x,y \in \{0,1\}^N} \weight(x,y)\; \inner{\psi_x^j}{\psi_y^j}.
\end{equation}
denote the weighted all-pairs inner product after $j$ iterations.
Initially, the total weight is $W_0 = W$.
After $T$ iterations, the total weight is $W_T =
\sum_{x,y \in \{0,1\}^N} \weight(x,y) \inner{\psi_x^T}{\psi_y^T}$.
If algorithm~$A$ is capable of distinguishing with certainty
between all pairs of oracles $(x,y) \in \{0,1\}^N \times \{0,1\}^N$ 
of nonzero weight, then $W_T=0$.
Conversely, if $W_T>0$ then there exists some pair of 
oracles $(x,y)$ with $\weight(x,y)>0$ between which algorithm~$A$
does not distinguish perfectly.

In~summary, initially all inner products are~1 and the
initial weight is therefore~$W$, whereas at the end of the 
computation all inner products are hopefully small and the
final weight~$W_T$ is therefore small.
If~the total weight can decrease by at most $\Delta$
by each query, we require at least $W/\Delta$ queries
to perfectly distinguish between all pairs of oracles
of nonzero weight.

\begin{theorem}\label{thm:general}
Let $f: S \rightarrow \{0,1\}^\emm$ be a given function where
$S \subseteq \{0,1\}^N$,
and let $\omega$ be a weight function for~$f$.
Let $A = (\op{U}\op{O})^T \op{U}$ 
be any quantum algorithm that computes~$f$
with error at most~$\epsilon \geq 0$ using $T$ queries.
Then 
\begin{equation}
T \geq \bigg(1-2\sqrt{\epsilon(1-\epsilon)}\bigg) \frac{W}{\Delta}
\end{equation}
where $W = \sum_{x,y \in \{0,1\}^N} \weight(x,y)$ denotes
the initial weight, and
$\Delta$ is an upper bound on $|W_{j} - W_{j+1}|$ for 
all $0\leq j< T$.
\end{theorem}

\begin{proof}
By~definition, $W_0 = W$, and by Lemma~\ref{lm:epsilon}, 
$|W_T| \leq 2\sqrt{\epsilon (1-\epsilon)} \smallspace W$.
Write
$W_0 - W_T = \sum_{j=0}^{T-1} (W_j - W_{j+1})$ as a telescoping
sum.
Then
$|W_0 -W_T| \leq \sum_{j=0}^{T-1} |W_j - W_{j+1}| \leq T \Delta$,
and the theorem follows.
\end{proof}

Our formulation of Theorem~\ref{thm:general} has been heavily
inspired by the general formulations used by Ambainis 
in~\cite{Ambainis2}.
In~\cite{BBBV}, 
Bennett, Bernstein, Brassard and Vazirani 
are interested in distinguishing
one unique oracle~$x'$ from all other oracles.
That is, for every pair of oracles 
$(x,y) \in \{0,1\}^N \times \{0,1\}^N$ 
of interest, we have~$x=x'$.
Ambainis~\cite{Ambainis2} removes this restriction, and
he also allows a non-uniform interest in different oracles
by weighting each oracle individually.
We~are also interested in distinguishing general pairs of oracles, 
but we discriminate our interest in each pair by weighting 
each \emph{pair of oracles} via weight function~$\weight$.
This discrimination is essential in our application to
ordered searching.

\section{Lower bound on ordered search}
\label{sec:search}

Searching ordered lists is a non-Boolean promise problem: 
the list is promised to be sorted, and the answer is an index, 
not a bit.
Formally, the set~$S$ of $N$ possible inputs consists of 
all $x \in \{0,1\}^N$ for which $x_{0}=0$ and 
$x_{i-1} \leq x_{i}$ for all $1\leq i<N$.
The search function $f:S \rightarrow \{0,1\}^\emm$ 
is defined by $f(x) = \min\{ 0 \leq i<N \mid x_i =1\}$,
where we identify the result $f(x)$ 
with its binary encoding as a bit-string of 
length $\emm = \lceil \log_2(N)\rceil$.

The classical query complexity of this problem is,
as is well known, exactly $\lceil \log_2(N)\rceil$.
The best known quantum algorithm for this problem
requires roughly $0.526 \log_2(N)$ queries and is 
due to Farhi, Goldstone, Gutmann and Sipser~\cite{FGGS2}.
The first lower bound of $\sqrt{\log_2(N)}/\log_2\log_2(N)$ 
was proved by Buhrman and de~Wolf~\cite{BdW} 
by an ingenious quantum reduction from the \textsc{or} problem.
Farhi, Goldstone, Gutmann and Sipser~\cite{FGGS1} 
improved this to $\log_2(N)/2\log_2\log_2(N)$, and 
Ambainis~\cite{Ambainis1} then recently proved the
previous best known lower bound of $\frac{1}{12} \log_2(N) - O(1)$.
In~\cite{FGGS1,Ambainis1}, they use, as we do here, 
an inner product argument along the lines of~\cite{BBBV}.

The first and essential step in our lower bound is to 
pick a good weight function~$\weight$ for~$f$.  We~choose 
\begin{equation}\label{eq:weight}
\weight(x,y) =
\begin{cases}
\frac{1}{f(y)-f(x)}
 &\text{if $(x,y) \in S \times S$ and $f(x)<f(y)$}\\
0  &\text{otherwise.}
\end{cases}
\end{equation}
That is, we use the inverse of the Hamming distance 
of $x$ and~$y$.
Intuitively, a weight function that (only) depends on the 
Hamming distance ought to be a good choice since it can put
most weight on pairs of oracles that are almost identical.

The initial weighted all-pairs inner product is
\begin{equation}
W_0 = \sum_{x,y \in \{0,1\}^N} \weight(x,y)
= \sum_{i=1}^{N-1} H_i = N H_N - N,
\end{equation}
where $H_i = \sum_{k=1}^{i} \frac{1}{k}$ denotes
the \nth{i} harmonic number.
Note that $\ln(N) < H_N < \ln(N)+1$ for all $N>1$.
Since any query can decrease the weighted all-pairs inner product 
by at most~$\pi N$, our main theorem follows by applying 
Theorem~\ref{thm:general}.

\begin{lemma}\label{lm:search}
For weight function~$\weight$ defined by 
Eq.~\ref{eq:weight}, we have that
\begin{equation*}
|W_j-W_{j+1}| \leq \pi N
\end{equation*}
for all $0 \leq j <T$.
\end{lemma}

\begin{theorem}[Main]
Any quantum algorithm for ordered searching that errs with
probability at most~$\epsilon \geq 0$ requires at least
\begin{equation}
\bigg(1-2\sqrt{\epsilon(1-\epsilon)}\bigg) 
\frac{1}{\pi}\tinyspace(H_N-1)
\end{equation}
queries to the oracle.
In~particular, any exact quantum algorithm requires more than
$\frac{1}{\pi}(\ln(N)-1)$ queries.
\end{theorem}

We~end this section by given our proof of Lemma~\ref{lm:search}.

\begin{proofof}{Lemma~\ref{lm:search}}
For any oracle $x \in \{0,1\}^N$, we will think of~$x$ as 
an infinite bit-string where $x_i=0$ for all $i\geq N$.
Operator~$\op{O}_x$ defined by Eq.~\ref{eq:oracle} is then given~by 
\begin{equation*}
\op{O}_x = \sum_{z \geq 0} 
     \sum_{i \geq 0} (-1)^{x_i}\ket{z;i}\bra{z;i}.
\end{equation*}
Let $\identity$ denote the identity operator.
For every $i\geq 0$, let
$\projection_i = \sum_{z\geq 0} \ket{z;i}\bra{z;i}$
denote the projection operator onto the subspace querying 
the \nth{i} oracle bit.  

Let $0 \leq j<T$.  By~definition 
\begin{align*}
W_j - W_{j+1} 
&= \sum_{x,y \in \{0,1\}^N} \weight(x,y)\; \inner{\psi_x^j}{\psi_y^j}
   - \sum_{x,y \in \{0,1\}^N} \weight(x,y)\; 
   \inner{\psi_x^{j+1}}{\psi_y^{j+1}}\\
&= \sum_{x,y \in \{0,1\}^N} \weight(x,y)\; 
   \tinner{\psi_x^j}{\identity - \op{O}_x^{-1} \op{O}_y}{\psi_y^j}\\
&= 2 \sum_{\smash[b]{x,y \in \{0,1\}^N}}
   \sum_{i : x_i \neq y_i\vphantom{\}^N}}
   \weight(x,y)\; \tinner{\psi_x^j}{\projection_i}{\psi_y^j}.
\end{align*}
For every $0 \leq a<N$ and $i \geq 0$, let 
$\beta_{a,i} = \projection_i \ket{\psi_x^j}$, where
$x \in \{0,1\}^N$ is such that $f(x)=a$ ('$a$' for 'answer').
Then
\begin{equation*}
W_j - W_{j+1}
= 2 \sum_{0 \leq a < b<N} \sum_{a \leq i<b}
  \frac{1}{b-a} \betastar_{a,i} \betastaralign_{b,i},
\end{equation*}
where $c^{*}$ denotes the complex conjugate of~$c \in \complex$.
Rewrite the above equation in terms of distances $d=b-a$,
\begin{equation*}
W_j - W_{j+1}
= 2 \sum_{d=1}^{N-1} \sum_{i=0}^{d-1} \frac{1}{d} 
  \Bigg(\sum_{a=0}^{N-d-1} \betastar_{a,a+i} 
  \betastaralign_{a+d,a+i}\Bigg).
\end{equation*}
For every $0\leq i<N-1$, let
\begin{equation*}
\gamma_i = \Bigg(\sum_{a=0}^{N-1} |\beta_{a,a+i}|^2\Bigg)^{1/2}
\;\quad \text{and } \;\quad
\delta_i = \Bigg(\sum_{a=0}^{N-1} |\beta_{a,a-i-1}|^2\Bigg)^{1/2}
\end{equation*}
denote the total mass that queries the oracle at 
$i$~index-positions above and below the leftmost~1.
By~the Cauchy--Schwarz inequality,
\begin{equation*}
|W_j - W_{j+1}|
\leq 2 \sum_{d=1}^{N-1} \sum_{i=0}^{d-1} \frac{1}{d} 
  \gamma_i \delta_{d-i-1}\tinyspace.
\end{equation*}

The right hand side is the written-out product of 3~matrices.
Let $\gamma = [\gamma_0,\ldots,\gamma_{N-2}]$ and 
$\delta = [\delta_0,\ldots,\delta_{N-2}]^t$, where $t$ denotes
transposition, and
let $M$ denote the $(N-1) \times (N-1)$ matrix with
entry $(k,\ell)$ defined~by
\begin{equation*}
(M)_{(k,\ell)} = 
\begin{cases} 
\frac{1}{k+\ell+1} & \text{if $k+\ell<N-1$}\\
0 & \text{otherwise}
\end{cases}
\end{equation*}
for all $0 \leq k,\ell <N-1$.
Then 
\begin{equation*}
|W_j - W_{j+1}| \smallspace\leq\smallspace 2\gamma M \delta
\smallspace\leq\smallspace 
2 \|\gamma\|_2 \cdot \|M\|_2 \cdot \|\delta\|_2
\end{equation*}
where $\|K\|_2 = \max\{ \|Kv\|_2 : \|v\|_2=1\}$ 
denotes the induced matrix norm.

Since $\|\gamma\|_2^2 + \|\delta\|_2^2 \leq 
\sum_{a=0}^{N-1} \sum_{i\geq0} |\beta_{a,i}|^2 \leq N$,
we have that 
$\|\gamma\|_2 \smallspace \|\delta\|_2 \leq \frac{1}{2}N$.
Matrix~$M$ is a Hankel matrix, and its norm is upper 
bounded by the norm of the $(N-1) \times (N-1)$ Hilbert 
matrix $M'$ defined by $(M')_{(k,\ell)} = \frac{1}{k+\ell+1}$
for all $0 \leq k,\ell <N-1$.
The norm of any Hilbert matrix is upper bounded 
by~$\pi$ (see for example~\cite{Choi} for a neat argument),
and hence
$|W_j-W_{j+1}| \leq \pi N$.
\end{proofof}

\section{A~{\boldmath$\log_3(N)$} algorithm for ordered search}
\label{sec:upper}

In~this section, we sketch the main elements in our 
new quantum algorithm for ordered searching.
First, we give some general remarks, outline
the main parts and discuss the intuition behind the algorithm.
We then, in Subsection~\ref{subsec:analysis}, 
comment on parts of the analysis of its complexity.  
Our algorithm is exact and requires only 
$\log_3(N) + O(1)$ queries to the oracle.

Our algorithm 
bear many resemblances to the classical binary search algorithm.
All unitary transformations are ``nice''
in the sense that the algorithm only uses 
$\op{XOR}$ operations (also referred to as controlled-NOT)
and rotations by multiples of angle~$\frac{2\pi}{16}$.
The requirements to the resolution of the rotations 
in the algorithm are thus moderate.

In~the algorithm, we use a very important idea of 
\textit{explicitly known bits of the oracle}.
If~the algorithm can increase the number of 
explicitly known bits by a factor of~$F$ by each query,
then we obtain a complexity of $\log_F(N)+O(1)$. 

Consider an application of the classical binary search algorithm
on an oracle $x \in \{0,1\}^8$ of length $N=8$.
By~definition of the ordered search problem 
we are promised that $x_8=1$ and that $x_i \leq x_{i+1}$ for
all $i=1,\ldots,7$.
We~say that a bit of the oracle is \emph{explicitly known} 
if the classical algorithm can output the value of that 
bit with certainty for all possible input oracles.
Initially, only the bit $x_8$ is explicitly known
(since we are promised that $x_8=1$).
After the first query, the bit $x_4$ is also explicitly known,
and after the second query, the bits $\{x_2, x_4, x_6, x_8\}$ are
explicitly known, since independently of which oracle the 
algorithm is given, it knows the value of these four bits 
with certainty.
Finally, after the third and last query, all 8~bits are explicitly 
known. 

Since it is important for what follows, please do not confuse 
bits that are \emph{explicitly known} with bits who's value 
can be deduced only for \emph{some} oracles.
As~an example, suppose we run the classical binary search 
algorithm on oracle~$x$ for which $x_5=0$ and $x_6=1$.
After the second query, the classical binary search algorithm 
knows that $x_4=0$ and that $x_6=1$.
{From} this, it can deduce the values of all 8 bits, but $x_5$.
However, it is only the values of the four bits 
$\{x_2, x_4, x_6, x_8\}$ that can be deduced no matter
what oracle is given as input, and thus it is only these 
four bits we consider explicitly known.
The concept of explicitly known bits for the classical binary search
algorithm is very simple: for an input of length $N=2^n$, after
$j$ queries, there are exactly $2^j$ bits explicitly known,
these being $\{x_{N/2^j}, x_{2N/2^j}, \ldots, x_{N}\}$.
The concept of explicitly known bits for our quantum algorithm is 
almost as simple and it is defined below.

Our quantum algorithm is, in many senses, simply to run 
many copies of (variations of) the classical binary search algorithm 
in quantum parallel.  
Every basis state in every one of the superpositions we ever obtain,
has a natural classical interpretation as being the state 
of a classical computer running some classical search algorithm.  
Therefore, we shall often refer to basis states of a superposition 
as~\emph{computers}.  The point being that every basis state
can be thought of as the state of a classical computer in the
middle of searching.
By~making queries in quantum parallel we achieve that some of the
computers know more about where in the list the shift 
from \mbox{0's} to \mbox{1's}~is. 
These computers then communicate their knowledge to the rest
of the computers in the superposition.

With this, we can define we mean by explicitly known bits
for our quantum algorithm:
\emph{We~say that a {computer} in a superposition
explicitly knows a bit if it knows the value of that bit 
regardless of which oracle the algorithm is given as input.}
{To}~illustrate what we hereby mean, consider the previous example
in which we are given an oracle $x \in \{0,1\}^8$ of length
$N=8$ with $x_5=0$ and $x_6=1$.

Suppose that we have superposition of 4~computers.
We~want to find the shift from \mbox{0's} to \mbox{1's} 
in the oracle. 
Suppose that 2~of the computers explicitly know 4~bits 
(the bits 2, 4, 6, and~8), 
that 1~computer explicitly knows 2~bits (bits 4 and~8) 
and that 1~computer explicitly knows only 1~bit (bit~8). 
The following sequence of operations illustrates how we, 
using only a single query, 
combine the 4~computers so they all know 8~bits. 
\begin{gather*}
\begin{array}{r}
\ket{0}\iket{1}{8}\\
\ket{1}\iket{5}{8}\\
\sqrt{2}\ket{1}\iket{5}{6}
\end{array}
\overset{\op{O}'}{\longrightarrow}
\begin{array}{r}
\ket{x_4}\iket{1}{8}\\
(-1)^{x_6}\ket{1}\iket{5}{8}\\
\sqrt{2}(-1)^{x_5}\ket{1}\iket{5}{6}
\end{array}
\overset{\op{V}^{(8)}}{\longrightarrow}
\begin{array}{r}
\ket{0}\iket{5}{8}\\
(-1)^{x_6}\ket{0}\iket{5}{8}\\
\sqrt{2}(-1)^{x_5}\ket{1}\iket{5}{6}
\end{array} 
\overset{\op{U}^{(4)}}{\longrightarrow}
\\[.4cm]  
\begin{array}{r}
\sqrt{2}\ket{x_6}\iket{5}{8}\\
\sqrt{2}(-1)^{x_5}\ket{1}\iket{5}{6}
\end{array}
\overset{\op{V}^{(4)}}{\longrightarrow}
\begin{array}{r}
\sqrt{2}\ket{0}\iket{5}{6}\\
\sqrt{2}(-1)^{x_5}\ket{1}\iket{5}{6}
\end{array}
\overset{\op{U}^{(2)}}{\longrightarrow}
\begin{array}{r}
\sqrt{4}\ket{x_5}\iket{5}{6}
\end{array}
\overset{\op{V}^{(2)}}{\longrightarrow}
\begin{array}{r}
\sqrt{4}\ket{0}\iket{6}{6}.
\end{array}
\end{gather*}
We~have here (without loss of generality) 
assumed that the query operator~$\op{O}'$ is such 
that all computers but the one that knows the least have
their answers given in the phase of their amplitudes. 
The computer that knows the least is given the answer 
directly into its leftmost qubit. 

\begin{figure}
\begin{center}
\epsfig{file=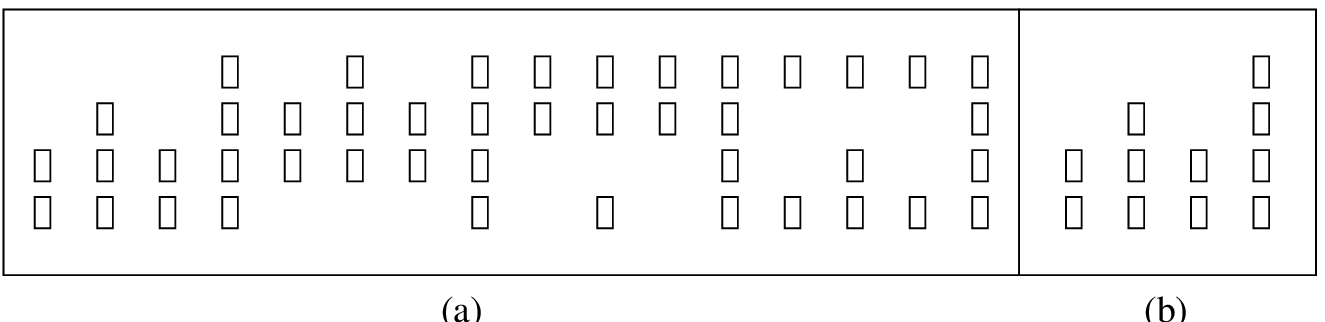}
\caption{
(a) Consider an oracle of size $N=32$.
The graphical representation of how 4 computers each arranges
their 11~explicitly known bits of the oracle so that using
only 1~query, all 4 computers learn all 32 bits. 
Each row corresponds to 1~computer, 
and each small block corresponds to 1~explicitly known 
bit of the oracle.
For instance, the first computer explicitly knows the 
11~bits $\{x_8, x_{12}, x_{16}, x_{18}, \ldots, x_{32}\}$.
Using 4~computers gives 4~symmetrical sublists. 
Figure~(b) illustrates one of these symmetrical sublists.}
\label{f2}
\end{center}
\end{figure}
{To}~be able to combine the knowledge from several computers we have 
them distribute their explicitly known bits in a certain way. 
In Figure~\ref{f2}a it is shown how we for each one of 4 computers 
distribute 11 explicitly known bits 
so that after 1~iteration all 4~computers explicitly know 32~bits.
The distribution is so that we divide the list into 4 sublists
of equal size, where we for each sublist distribute 11 bits 
as shown in Figure~\ref{f2}b. 
Similarly, if we have $2^n$ computers, we divide the list 
into $2^n$ sublists of equal size.
In~general, the more computers we use, the more sublists 
we have, and the more explicitly known bits each computer 
needs to have.

\subsection{The Algorithm}
\label{subsec:algorithm}
In~each iteration of the algorithm, we utilize 2~operators,
\up{\op{U}}{s} and \up{\op{V}}{s}, where 
\up{\op{U}}{s} is used to combine 2~computers 
and \up{\op{V}}{s} makes a computer ready for 
the next \up{\op{U}}{s} operator, which is defined as follows.
\begin{equation}
\op{U}^{(s)} :\; \ket{b}\iket{\uell}{h} \;\longmapsto\; \left \{ 
    \begin{array}{ll}
      \frac{1}{\sqrt{2}}(\ket{0} + (-1)^b \ket{1})\iket{\uell}{h} &
      \textrm{ if } s=h-\uell+1 \\
      \ket{b}\iket{\uell}{h} & \textrm{ otherwise.}
    \end{array}
\right.
\label{trick}
\end{equation}
Here $b$ is a single bit, $s$~is a power of~2, and 
$\uell$ and~$h$ are the endpoints of an interval.

We use the operator defined by Eq.~\ref{trick} as in the following
\begin{equation}
\alpha \ket{0}\ket{\up{I}{s}} + \alpha
  (-1)^{\up{x}{\frac{s}{2}}}\ket{1}\ket{\up{I}{s}}
  + \beta \ket{\Phi} \quad
  \overset{\up{\op{U}}{s}}{\longrightarrow} 
  \quad  \sqrt{2}\alpha \ket{\up{x}{\frac{s}{2}}} 
  \ket{\up{I}{\frac{s}{2}}}+ \beta \ket{\Phi}
\end{equation}
where \up{I}{s} is an interval, \up{x}{\frac{s}{2}} is the bit at
the middle of the interval, $\alpha$ and $\beta$ are real-valued
amplitudes satisfying 
$2\alpha^2+\beta^2=1$, and \ket{\Phi} is a superposition 
orthogonal to \ket{0}\ket{\up{I}{s}} and \ket{1}\ket{\up{I}{s}}.

The operators $\op{V}^{(s)}$ are defined~by
\begin{equation}
\op{V}^{(s)} :\; \ket{b}\iket{\uell}{h} \;\longmapsto \; \left \{ 
    \begin{array}{ll}
      \ket{0}\iket{\uell}{(\uell+\frac{h-\uell+1}{2}-1)} & ~ \textrm{if } s=h-\uell+1
      \land b=1 \\
      \ket{0}\iket{(\uell+\frac{h-\uell+1}{2})}{h} & ~ \textrm{if } s=h-\uell+1
      \land b=0 \\
      \ket{b}\iket{\uell}{h} & ~ \textrm{otherwise.}
    \end{array}
\right .
\label{def_of_V}
\end{equation}
We~always use operator $\op{V}^{(s)}$ as in the following,
\begin{equation}
\ket{\up{x}{\frac{s}{2}}}\ket{\up{I}{s}} + \beta \ket{\Phi} \quad
\overset{\up{\op{V}}{s}}{\longrightarrow} \quad
\ket{0}\ket{\up{I}{\frac{s}{2}}} + \beta \ket{\Phi}.
\end{equation}
Each operator $\op{V}^{(s)}$ takes a interval and the value of the 
bit~$x^{(\frac{s}{2})}$ 
in the middle of that interval, and chooses the upper 
part of that interval if $\up{x}{\frac{s}{2}}=0$, 
and chooses the lower part if $\up{x}{\frac{s}{2}}=1$.

The operators are applied to the state in the following fashion
\begin{equation}
\op{V}^{(2)} \op{U}^{(2)} \cdots 
  \op{U}^{(N/2)} \op{V}^{(N)} \op{U}^{(N)} \op{V}^{(2N)} \op{O}'.
\label{iteration}
\end{equation}

Assume that we have $r=2^n$ computers which have distributed their
explicitly known bits in a way such that we obtain $r$
sublists. 
The shift from 0s to 1s in the list is in one and only one of 
these sublists.  Consider the sublist containing the shift.
First all computers make a query to the oracle in quantum parallel.
Then we combine the computers using the 
2~operators $\op{U}^{(s)}$ and~$\op{V}^{(s)}$. 
These operators are applied $\log(r)$ times after a query.
After applying this sequence of operations all computers
know where in the sublist the shift is.
As~we have seen $\op{V}^{(s)}$ changes
the name of the computers so that they can be combined. 
Operator~$\op{U}^{(s)}$ combine 2 (sets of) computers. 

Suppose that after the application of 
operator \up{\op{U}}{s'} with, say,
$s'=\frac{r}{2^k}$ the state is
\begin{displaymath}
(\sqrt{2})^k \ket{\up{x}{\frac{s'}{2}}}\ket{\up{I}{s'}} 
+ (\sqrt{2})^k (-1)^{\up{x}{\frac{s'}{4}}} 
  \ket{1}\ket{\up{I}{\frac{s'}{2}}} + \beta \ket{\Phi}.
\end{displaymath}
We now apply the operator \up{\op{V}}{s'} which gives us the state
\begin{displaymath}
(\sqrt{2})^{k} \ket{0}\ket{\up{I}{\frac{s'}{2}}} + (\sqrt{2})^k
(-1)^{\up{x}{\frac{s'}{4}}} \ket{1}\ket{\up{I}{\frac{s'}{2}}} 
+ \beta \ket{\Phi},
\end{displaymath}
and after the application of operator \up{\op{U}}{\frac{s'}{2}} 
we have
\begin{displaymath}
(\sqrt{2})^{k+1} \ket{\up{x}{\frac{s'}{4}}}\ket{\up{I}{\frac{s'}{2}}} 
  + \beta\ket{\Phi}.
\end{displaymath}
{From} this we can verify that all computers obtain the explicit
knowledge of $2r$ bits (in this sublist) after $\log(r)-1$ 
applications of \up{\op{U}}{s} and $\log(r)$ of \up{\op{V}}{s}. Since
the operators $\op{U}^{(s)}$ and $\op{V}^{(s)}$ are unitary it
follows that the operator given by Eq.~\ref{iteration} is unitary. 

To be able to apply the iteration again after another query,
one must make the above iteration in a superposition 
so that there now is a superposition of computers that have their 
explicitly known bits
distributed in a structure resembling figure~\ref{f2}. 
We explain how this is done, including other omitted parts,
in the final version of this paper.

\subsection{Elements of the analysis}
\label{subsec:analysis}
Our routine described above can be utilized to increase
to number of known bits by a factor of almost~3 by
each iteration.
Consider figure~\ref{f2}(a).  
In this figure, we use 4 computers, one corresponding to
each of the 4 rows.  Each of the 4 computers knows 11 bits
(1+2+4+4) before the iteration, whereas after the iteration, 
all of them know 32 bits.  Thus, using only 1~query, we 
increase the number of known bits from 11 to~32.

In~general, starting with $r=2^n$ computers, 
each of them explicitly knows $m$~bits of the oracle, 
where $m$ is given~by
\begin{equation}
m=1*1+ 1*2+2*4+\cdots + \frac{r}{2}*r = 1 + \sum ^{n-1} _ {k=0}
2^k*2^{k+1} = \frac{2}{3} 4^n +\frac{1}{3},
\end{equation}
after 1~iteration, all of the $r$ computers will know $m'=r*r*2$
bits.
Thus, with just 1~iteration we go from
having $m=\frac{1}{3}(2* 4^n +1)$ known bits to having 
$m'= r*r*2= 2*4^n$ known bits, an expansion factor of almost~3.

Given any value of~$m$ we can decompose it into 
the $\frac{1}{3}(2* 4^n +1)$ number system and obtain 
an expansion factor of almost~3.
For integer~$m$, write
\begin{equation*}
m = \sum ^a _{k=0} \alpha _k *  \frac{1}{3}(2* 4^k +1),
\label{a1}
\end{equation*}
where $\alpha _k \in \{0,1,2,3\}$ for $k=0,1,\ldots,a$,
with $a = \max \{ k \mid \alpha _k \ne 0 \}$.
Using 1~query, we expand $m$ known bits into $m'$ known 
bits where
\begin{equation*}
m' = \sum ^a _ {k=0}  \alpha _k *  2* 4^k.
\end{equation*}
We achieve an expansion factor of 
\begin{equation}
F = \frac{m'}{m} 
= \frac{\sum ^a _ {k=0}  \alpha _k *  2* 4^k}
 {\frac{1}{3} \sum ^a _ {k=0}  \alpha _k *  (2* 4^k + 1)} 
\geq \frac{3}{1+\frac{3(a+1)}{2*4^a}}
= 3 - O\bigg(\frac{\log(m)}{m}\bigg).
\end{equation}

\section{Concluding remarks}
\label{sec:conclusion}

The inner product of two quantum states is a measure for their 
distinguishability.  
For instance, two states can be distinguished with certainty 
if and only their inner product is~0.
In~this paper, we have introduced a weighted all-pairs inner product 
argument as a tool for proving lower bounds in the 
quantum black~box model.
We~have used this argument to give a better and also simpler 
lower bound of $\frac{1}{\pi}(\ln(N)-1)$ for quantum ordered searching.
It~seems to us that the possibility of using non-uniform 
weights is particular suitable when proving lower bounds 
for non-symmetric (possibly partial) functions.

We~have chosen here to use inner products which is only one 
of the many studied measures for distinguishability of states.
A~striking example of the limitations of using this measure
is given by Jozsa and Schlienz in~\cite{JS}.
In~\cite{Zalka}, Zalka uses a non-linear measure to prove 
the optimality of Grover's algorithm~\cite{Grover}.
\samepage{
Similarly, it might well be that utilizing some other 
(possibly non-linear) measure of distinguishability could 
be used to improve our lower bound.

Our algorithm for searching an ordered list with complexity 
$\log _3(N) +O(1)$ 
is based on classical binary search combined with}
\pagebreak[4]
communicating knowledge between computers or basis states. 
It~could be instructive to studying
if this view on quantum computation could be useful
in constructing quantum algorithms for other problems.  
Either in the sense if it is possible 
quantumly to speed up classical algorithms for other problems
(where amplitude amplification does not apply) 
or if it is possible in solutions for other problems
to take advantage of communication between computers.


\end{document}